\documentclass{iopart}
\input{epsf.sty}
\input{amssymb.sty}

\begin{document}

\hyphenation{extra-po-la-tion para-meters}

\title[Structural Relaxation and Mode Coupling in a Simple Liquid: Benzene]
{Structural Relaxation and Mode Coupling in a Simple Liquid:
Depolarized Light Scattering in Benzene}

\author{Sabine~Wiebel and Joachim~Wuttke \footnote{Corresponding author. 
                  Also at Siemens AG, ICN ON RD AT~1,
		  Hofmannstr.~51, 81359 M\"unchen, Germany.
 		  E-mail: {\tt jwuttke\makeatletter @\makeatother ph.tum.de} 
		  URL: {\tt http://}\linebreak[2]{\tt
		  www.e13.}\linebreak[2]{\tt physik.}\linebreak[2]{\tt
		  tu-}\linebreak[2]{\tt muenchen.}\linebreak[2]{\tt
		  de/Wuttke}}}

\address{Physik-Department E13, Technische Universit\"at M\"unchen, 
         85747 Garching, Germany}


\begin{abstract}
We have measured depolarized light scattering in liquid benzene over the whole
accessible temperature range and over four decades in frequency.
Between 40 and 180~GHz we find a susceptibility peak
due to structural relaxation.
This peak shows stretching and time-temperature scaling 
as known from $\alpha$ relaxation in glass-forming materials.
A simple mode-coupling model provides consistent fits
of the entire data set.
We conclude that structural relaxation in simple liquids 
and $\alpha$ relaxation in glass-forming materials are physically the same.
A deeper understanding of simple liquids is reached by applying concepts
that were originally developed 
in the context of glass-transition research.
\end{abstract}
\pacs{61.20.Lc,61.25.Em,78.35.+c,64.70.Pf}


\section{Motivation}


On short time scales, all liquids show solid-like elasticity.
This has been
impressively illustrated by Brillouin scattering of X-rays \cite{SeRK96}:
on a THz scale, sound propagates in water with almost the same speed as in ice,
more than twice as fast
as on the kHz or MHz scale of conventional ultrasonic measurements.
Such a cross-over goes along with a decay of structural correlations;
it is called {\it relaxation}, and more specifically $\alpha$ relaxation
when it leads from solid-like to liquid-like response
(correlations decaying to zero).

When a liquid can be supercooled far enough,
$\alpha$ relaxation becomes critically slow,
so that the material ultimately becomes a glass.
The dynamics of glass-forming liquids has been studied in great detail.
Stretching and time--temperature scaling have been identified 
as generic properties of $\alpha$ relaxation.
Additional scaling laws have been uncovered by a mode-coupling theory (MCT).
Originally proposed as a theory of the glass transition,
MCT is now generally recognized to offer
a unified description of microscopic and relaxational motion
at comparatively high temperatures 
where $\alpha$ relaxation occurs on a GHz scale.

In experiments undertaken in both the supercooled and the normal liquid phase,
$\alpha$ relaxation and mode-coupling dynamics are found to evolve continuously
across the melting point of the concurrent crystalline phase:
on a GHz to THz scale,
the molecular dynamics seems to be insensitive to
 whether the liquid's state is thermodynamically stable or not.
This leads us to the hypothesis
that the molecular dynamics in the normal liquid state will be insensitive to
 whether the liquid can be supercooled or not.
A pioneering paper on water \cite{SoHQ95b},
studies of metallic melts \cite{MeWP98a,MeBS99b},
and our own experiments on several molecular liquids
suggest that indeed
$\alpha$ relaxation and mode-coupling effects
also occur in liquids that cannot be supercooled into a glass.

For a more detailed test of our hypothesis,
we now study one such liquid in depth.
We choose benzene, which presents the following advantages:
(i) the molecule is structurally very simple and highly symmetric;
(ii) it is stiff on all relevant time or frequency scales \cite{x77};
(iii) many often studied glass formers are structurally related to benzene;
(iv) benzene is an excellent light scatterer.

To investigate the dynamics of benzene over a wide dynamic range,
we have used depolarized light scattering.
A neutron scattering study is currently under way
and will be published later;
preliminary results support the conclusions we draw from light-scattering.

\section{Context and Theory}

\subsection{\boldmath{$\alpha$} Relaxation}

In zeroth approximation, 
relaxation may be modelled by an exponential decay of correlations.
This ansatz goes back to Maxwell's theory of viscoelasticity;
it has been elaborated for dielectric response by Debye \cite{Deb29}.
The underlying physics is mean-field like:
one considers thermal motion of an individual molecule
upon which all the other molecules exert just a constant friction.

However, in glass-forming liquids
$\alpha$ relaxation is found to be {\it stretched}:
it is spread much more 
than an exponential decay in time
or a Lorentz line (in our context: a Debye resonance) in frequency.
Popular fit functions assume a fractional time or frequency dependence,
as in Kohlrausch's stretched exponential 
\begin{equation}\label{Ekww}
  \Phi_{\rm K} (t) \propto \exp{[-(t/\tau )^{\beta}]}\,,
\end{equation}
or in the Cole--Davidson susceptibility
\begin{equation}\label{Ecd}
   \chi_{\rm CD} (\nu) \propto [(1 + i2\pi \nu \tau)]^{-\beta}\,.
\end{equation}
While the relaxation time $\tau$ depend strongly on the temperature~$T$,
the exponents $\beta_{\rm K}$ or $\beta_{\rm CD}$ vary only weakly.
This can be seen as a consequence of {\it time-temperature scaling}:
in a good first approximation,
$\alpha$ relaxation has the form
\begin{equation}\label{Eama}
   \Phi (t;T) \simeq \hat{\Phi}(t/\tau(T))\,.
\end{equation}
Towards high temperatures, this scaling law has an obvious limitation:
$\alpha$ relaxation can never become faster than the
temperature-independent microscopic modes.
Indeed, on heating glass-forming materials towards the boiling point
the curves $\log \tau$ {\it vs.\/} $T$ become flatter and flatter,
so that $\tau$ approaches, 
but never reaches the intrinsic time scale of microscopic motion \cite{Ang97b}.
Depolarized light scattering in molecular liquids could clearly resolve
an $\alpha$ peak and confirm its scaling
up to temperatures far in the normal liquid phase \cite{CuLD97b,WiSR00b}.

\subsection{Relaxation in Simple Liquids}

Historically, relaxation in simple liquids (on a THz scale),
and $\alpha$ relaxation in highly viscous liquids (originally measured
mostly on Hz to MHz scales)
have long been seen as two genuinely different processes \cite{x81}.
In the most simple monatomic liquids
like argon or sodium,
on which theory-oriented textbooks 
 \cite{BoYi80,HaDo86,BaZo94} concentrate,
characteristic relaxation times are of the order of 
$10^{-13}$ to $10^{-12}$~s,
which is not much longer than the mean time between collisions \cite{x78}.
Under such circumstances relaxation is closely mingled with
microscopic motion,
and it is impossible to obtain isolated experimental information
on relaxation alone.
On the other hand,
results of scattering experiments and molecular dynamics simulations
cannot be understood without taking into account relaxation.
Therefore, experimental data are usually fitted by
theoretical expressions that contain
memory kernels built upon an {\it ad-hoc} model of relaxation.

Such fits, however, are rather insensitive
to the functional form of the memory kernel,
and therefore
one seldom went beyond assuming simple exponential relaxation.
In some cases, when fits were judged unsatisfactory,
a sum of two exponentials was used \cite{x80};
this approach, though admittedly arbitrary,
has recently been revived in the analysis of X--ray Brillouin scattering
on liquid metals \cite{ScBR00a,ScBR00b,ScBR02}.

In the investigation of glass-forming liquids 
this double exponential approach has long been overcome
by formul\ae\ like (\ref{Ekww}) or~(\ref{Ecd}).
Today, after $\alpha$ relaxation has been observed across the melting point
and up to a GHz to THz scale, 
it is no longer justified to consider relaxation in glass-forming materials
and relaxation in simple liquids as two different physical processes.
So we are led to suspect that the stretching and scaling properties
of $\alpha$ relaxation hold in principle even in argon and sodium,
although an experimental verification will be extremely difficult \cite{x84}.

For the time being we prefer to investigate
a molecular, non glass-forming liquid,
benzene, which in a sense is intermediate between monatomics on the one side
and molecular glass formers on the other side:
the benzene molecule is small and highly symmetric,
in contrast to glass-forming liquids which necessarily 
have a more complicated structure as to prevent crystallisation.
Yet
we will find structural relaxation in benzene to be slow enough \cite{x81}
to allow for a direct, unambiguous observation of its stretching.

\subsection{Mode-Coupling Theory}\label{Smct}

As mentioned in the beginning,
mode-coupling theory (MCT)
\cite{Got91,GoSj92}
provides a unified description of 
$\alpha$ relaxation and low-frequency vibrations.
It is a microscopic theory,
formulated as function of wavenumber~$q$ and time~$t$,
and built upon the static structure factor $S(q)$
and the density correlation function $\Phi_q(t)=S(q,t)/S(q)$.
It starts with the formally exact equation of motion
\begin{equation}\label{Eqmo}
  0=\Omega_q^{-2}\ddot{\Phi}_q(t) + 
    \Phi_q(t) + \int_0^t\!{\rm d} t'\,M_q(t-t')\dot{\Phi}_q(t')\,.
\end{equation}
The memory kernel $M_q(t)$ contains fast and slow fluctuations,
   $M_q(t) \simeq M_q'(t) + m_q(t)$.
The fast component, approximated as $M_q'(t)\simeq\gamma_q\delta(t)$,
can be shown to be irrelevant for the long-time behaviour.

The basic idea of MCT is now to expand
the slow fluctuations $m_q(t)$ in polynomials of density fluctuations,
and then to factorize all terms into pair correlations.
In lowest order one obtains the bilinear functional
\begin{equation}\label{Eker}
   m_q\{\Phi(t)\} = \sum_{{\bf p}+{\bf k}={\bf q}}
                V_{{\bf q}{\bf p}{\bf k}}
                \Phi_p(t)\,\Phi_k(t)
\,.\end{equation}
The coupling coefficients $V_{{\bf q}{\bf p}{\bf k}}$
can be derived from the static structure factor $S(q)$;
as $S(q)$,
they vary slowly with state variables like temperature~$T$ or pressure~$P$.

In this way the dynamics is completely determined
by a closed set of integro-differential equations.
Depending on the numeric values of $V_{{\bf q}{\bf p}{\bf k}}$,
the $\Phi_q(t)$ either decay to zero or arrest at finite values.
The border line~$T_{\rm c}(P)$
which separates these two cases
has been called the {\em ideal glass transition}.

For $T<T_{\rm c}$, the density correlations arrest
at a finite Debye-Waller factor $\Phi_q(t\!\to\!\infty) = f_q(T)$,
as expected for a glass.
On the liquid side, with $T>T_{\rm c}$,
the $\Phi_q(t)$ slow down on approaching a plateau $f_q(T)$,
but ultimately they decay to $\Phi_q(t\!\to\!\infty) = 0$
in a process which is easily identified as $\alpha$ relaxation.
Since the derivatives $\ddot{\Phi}_q(t)$ become negligible at long times,
Eq.~(\ref{Eqmo}) immediately reproduces 
the time-temperature scaling (\ref{Eama}),
with the corollary that 
the line shape of $\hat{\Phi}_q$ may vary with~$q$. 
To first order, 
solutions of MCT equations are consistent 
with the Kohlrausch asymptote~(\ref{Ekww}).

On cooling towards $T_{\rm c}$, $\alpha$ relaxation times are predicted
to diverge with a fractional power law in $T-T_{\rm c}$.
A comparison with measured relaxation times and viscosities 
\cite{TaKB86} shows
that such a divergence does {\em not} correctly describe
the glass transition \cite{x70}.
Instead, $T_{\rm c}$ is found to describe a cross-over
that is typically located 15--20~\% {\em above}
the conventional glass transition temperature $T_{\rm g}$:
while the density--density coupling of Eq.~(\ref{Eker})
becomes ineffective at $T_{\rm c}$,
other transport mechanisms, not covered by the theory,
remain active at lower temperatures.
This interpretation of $T_{\rm c}$ is 
supported by various other experimental indications of a cross-over
\cite{x74,x75,x76}.

In liquids to which MCT has been applied in the past,
the cross-over occurs at shear viscosities of the order of
$10^{1}$ to $10^{3}$~Poise \cite{TaKB86}.
We note that this is much closer to the viscosity of,
say, water at room temperature (about $ 10^{-2}$~Poise)
than to the glass transition 
(which, according to wide-spread convention, occurs at
$10^{13}$~Poise).
We note also that the dynamic predictions of MCT are expected to work best
not in the immediate vicinity of $T_{\rm c}$,
but at somewhat higher temperatures
where the concurrence of low-temperature transport processes can be neglected.

This suggests that MCT should be tested as a theory
that describes the dynamics of glass-forming liquids
at rather low viscosities, far above the glass transition, 
in a slightly supercooled or thermodynamically stable state.
In the present work we want to show that the restriction
to {\em glass-forming} liquids can be omitted altogether.

\subsection{Applying Mode Coupling to Real-Life Liquids}

Taken literally, Eqs.~(\ref{Eqmo}) and (\ref{Eker}) assume 
a liquid composed of identical, spherical, and stiff particles:
only in this case all interactions between the particles
can be derived from $S(q)$.
Extending the theory to mixtures is straightforward
\cite{BoTh87,BoKa95},
but including orientational and innermolecular degrees of freedom
poses extreme difficulties:
the notational and calculational efforts required by models so simple as
a liquid made of linear molecules \cite{ScSc97,FaLS99b},
or a dilute solution of linear molecules in spheres \cite{FrFG97c,GoSV00b}
are intimidating.
Thus, a MCT of molecular liquids is presently not available.

On the other hand,
the time and temperature dependence of MCT solutions 
is insensitive to most of the structural information
hidden in $V_{{\bf q}{\bf p}{\bf k}}$.
Taking into account orientational or innermolecular degrees of freedom
may lead to new classes of solutions,
but at least in some types of molecular liquids
the fundamental mathematical structure 
of equations (\ref{Eqmo}) and (\ref{Eker}) 
will remain dominant.

In such cases,
MCT solutions can be characterized by quite few parameters.
Close to $T_{\rm c}$, the 
analytical expansions of $\Phi_q(t)-f_q$ depends in lowest order on
just one nontrivial line shape parameter~$\lambda$.
Complete time correlation functions can be generated by numeric solutions
of very simple MCT models:

In the minimal $F_{12}$ model \cite{Got84},
just one correlator $\Phi(t)$ and two coupling coefficients in
\begin{equation}\label{Ephi}
  m(t)=  v_1 \Phi (t) + v_2 [\Phi (t)]^2
\end{equation}
are sufficient
to obtain relaxational stretching and the ideal glass transition.
With just one more correlator, $\Phi_{\rm s}(t)$, 
one can generate spectra with 
arbitrary $\alpha$ relaxation strengths $f_q^{\rm s}$:
a bilinear memory kernel \cite{BeGS84,Sjo86,x67}
\begin{equation}\label{Ephs}
  m_{\rm s}(t)=  v_{\rm s} \Phi (t) \Phi_{\rm s}(t) 
\end{equation}
couples $\Phi_{\rm s}(t)$ to $\Phi (t)$,
whereas $\Phi (t)$ does not depend on $\Phi_{\rm s}(t)$.
 
The so defined two-correlator $F_{12}$ model
is a physically meaningful tool for fitting experimental data.
As such it has already been successfully employed in several studies
of glass-forming liquids \cite{AlKr95,KrAK97,FrGM97,GoVo00};
the theoretical background is explained especially in Ref.~\cite{GoVo00}.
A detailed numeric study has confirmed the stability of such fits
\cite{KrAl02}.

\section{Light Scattering Measurements}

Benzene ($T_{\rm m}=279$~K, $T_{\rm b}=353$~K)
was bought from Sigma Aldrich
(99.9\% puriss.\ p.a.), unpacked under inert gas and sealed
into a Duran cuvette.
To our surprise,
the sample could be supercooled to 258~K 
where it remained liquid for several hours.
Data were taken at seven temperatures between 258 and 352~K.

Light scattering experiments were performed using
a grating double monochromator U\,1000
and a six-pass Sandercock-Fabry-Perot tandem interferometer.
In order to achieve stable operation at maximum resolution,
both instruments are placed in insulating housings
with active temperature control.
The optics around the interferometer has been modified 
as described previously \cite{WuOG00a,GoLW01a}.
Depending on the spectral range,
the interferometer is used in series with an interference filter
of either 150 or 1000$~$GHz bandwidth 
that suppresses higher-order
transmission leaks of the tandem interferometer
\cite{SuWN98,GaSP99,BaLS99} below 3\% or better. 
The filters are maintained at constant temperature.
To account for any drift, 
the instrument function is
redetermined periodically by automatic white-light scans.

In the present experiment, 
the slits of the monochromator are set to 30 -- 60 -- 60 -- 30~$\mu$m,
resulting in a resolution (fwhm) of 7.5~GHz;
data are only used above 200~GHz.
The interferometer is operated
with mirror spacings of 0.4, 2.8, and 16.3~mm,
corresponding to free spectral ranges of 375, 54, and 9.2~GHz;
some additional data were taken with 7.5~mm (20~GHz).

\begin{figure}
\epsfxsize=0.75\textwidth\centerline{\epsfbox{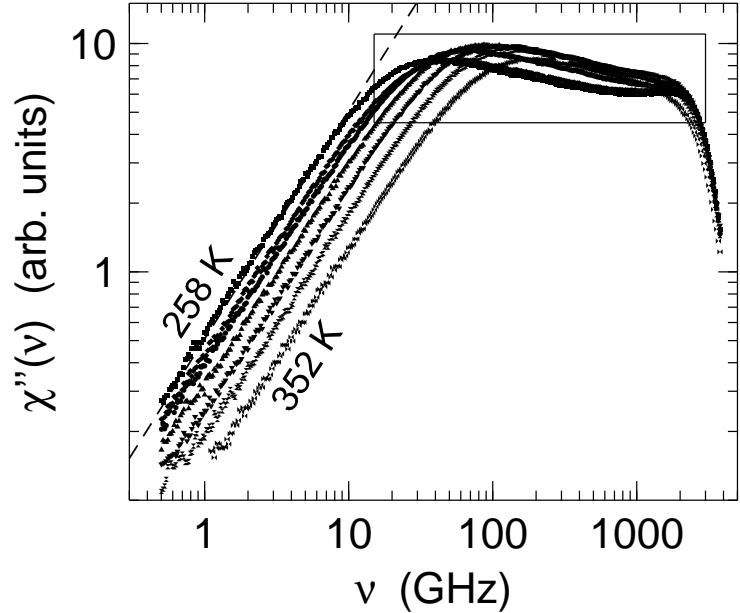}}\medskip
 \caption{Susceptibilities of benzene 
measured by depolarized light scattering 
at temperatures $T = $258, 268, 279, 293, 310, 330 and 352~K. 
In the low frequency wing, 
one finds the white-noise slope $\nu^1$,
indicated by the dashed line.
The nontrivial relaxational and microscopic dynamics 
inside the rectangular frame
is shown on an enlarged scale in Fig.~\ref{Fsusfit}.}
  \label{Fsus}
\end{figure}

On both spectrometers, 
a near-backscattering geometry ($172^{\circ}$) 
is used to minimize scattering from transverse modes.
In depolarized (HV) interferometer measurements,
the usual leakage from the acoustic modes is seen;
these lines are about a 100 times weaker than in polarized (VV) scattering,
but still up to about 3 times stronger than the continuous HV spectrum.
Subtracting separately measured VV spectra \cite{x85},
we could completely remove the contamination from the HV data.

Intensity calibration is always a problem in light scattering.
Best results were obtained by matching all data
to the middle (54~GHz) spectral range of the interferometer.
In this range intensities are reproduced
after a full temperature cycle within about 3~\%.
The temperature-dependent intensity mismatch of other spectral ranges
is higher and attains up to 20~\%.
Part of the problem may be due to distortions 
of the optical paths within the cryostat,
which are particularly severe when the spectrometer is operated with
small entrance opening.
In the present study,
intensities can also be estimated from a fit
which is normalized by construction
(see Figs.~\ref{Fnorm} below).

Finally, the spectra are multiplied with the detailed-balance factor,
\begin{equation}\label{Edetbal}
   \tilde{I}(\nu) = I(\nu) \exp(-h\nu/2k_{\rm B}T)
\,,
\end{equation}
averaged over energy-gain and energy-loss side,
and converted from intensity to susceptibility 
\begin{equation}\label{Echils}
\chi''_{\rm ls}(\nu)=\tilde{I}(\nu)/\tilde{n}(\nu)
\end{equation}
 with the symmetrized Bose factor 
\begin{equation}\label{Ebose}
  \tilde{n}(\nu) =
  {1\over \exp(h\nu/2k_{\rm B}T) - \exp(-h\nu/2k_{\rm B}T) }
\,.
\end{equation}
In other molecular liquids,
comparison with neutron scattering 
 \cite{WuOG00a,WuHL94,ToPD96,WuSH98,ShTB00,AoLD97}
has shown that depolarized light scattering 
yields at least qualitatively a good representation 
of the dynamic susceptibility,
and therefore we will interpret $\chi''_{\rm ls}(\nu)$
in very much the same way 
as a susceptibility from incoherent neutron scattering.

\begin{figure}
\epsfxsize=0.75\textwidth\centerline{\epsfbox{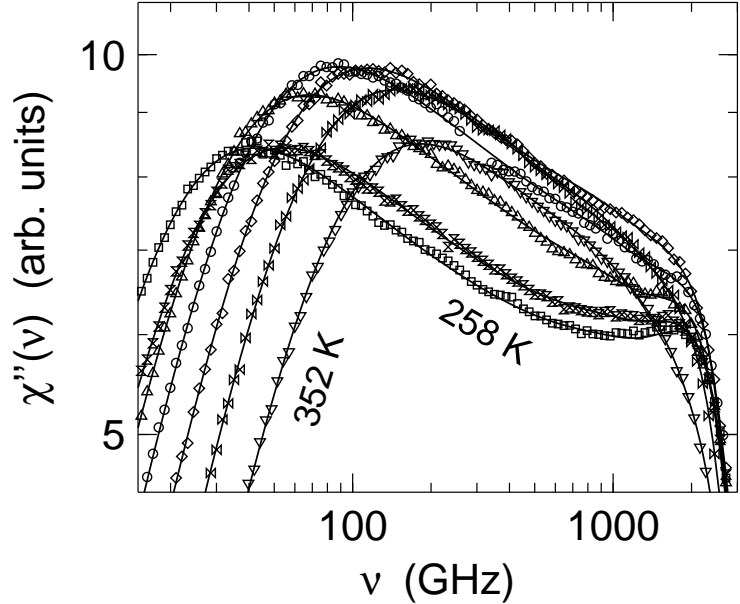}}\medskip
 \caption{Enlargement of 
the intermediate frequency region of Fig.~\protect\ref{Fsus}.
Structural $\alpha$ relaxation leads to a peak between 40 and 180~GHz.
The shoulder at about 2~THz is associated with microscopic ballistic motion.
The flat cross-over between these two peaks 
is a signature of mode-coupling dynamics:
the data cannot be explained as a simple superposition of $\alpha$ relaxation
and harmonic short-time motion.
Solid lines are fits with the mode-coupling two-correlator $F_{12}$ model
(see Fig.~\protect\ref{Fnorm},
Sects.~\protect\ref{Smct} and \protect\ref{Sexpmct} 
and Table~\protect\ref{Tf12}).}
\label{Fsusfit}
\end{figure}

\section{Data and Analysis}

\subsection{Susceptibilities on Logarithmic Scales}\label{Sraw}

Fig.~\ref{Fsus} shows susceptibilities from depolarized light scattering
for seven temperatures between 258 and 352~K.
In studies of glass-forming liquids,
measuring susceptibilities over several decades
and representing them on double logarithmic scales
were decisive steps in detecting nontrivial, 
stretched relaxation \cite{TaLC91a}.
In the case of benzene,
the same procedure, on the same absolute frequency scale,
is less rewarding:
too much of Fig.~\ref{Fsus} is filled by an uninformative 
$\nu^1$ white-noise wing.

Therefore we show the nontrivial part of our data in
Fig.~\ref{Fsusfit} on an enlarged scale:
the strongly temperature-dependent dynamics between 15 GHz and 3 THz.
With increasing frequency,
the $\chi''(\nu)$ begin to deviate 
from the white-noise limit $\chi''\propto\nu^1$,
reaching a maximum between 40 and 180~GHz 
which we will ascribe to structural $\alpha$~relaxation.
A comparatively flat region leads over to a shoulder
at about 2 THz, 
above which the susceptibilities strongly decrease.
Above 5 THz we find an extended gap;
innermolecular excitations 
are only expected above 12~THz \cite{x77}.

\begin{figure}
\epsfxsize=0.8\textwidth\centerline{\epsfbox{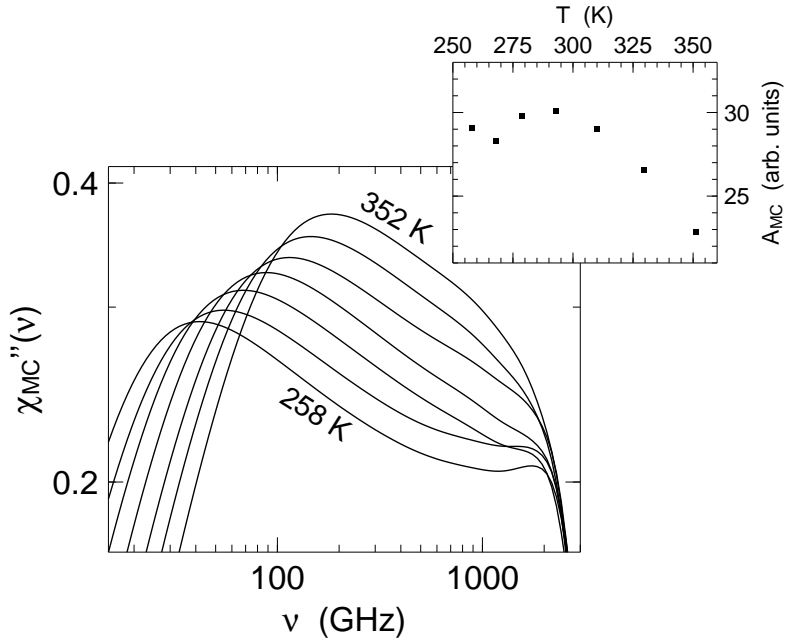}}\medskip
 \caption{The curves in the main figure are identical
to the mode-coupling fits of Fig.~\protect\ref{Fsusfit},
except for being shown in absolute units,
fulfilling  the $\chi''/\nu$ sum rule.
The inset shows the amplitudes $A_{\rm MC}(T)$ by which these curves
had to be multiplied in order to fit the experimental data;
these amplitudes essentially represent the Pockels coefficient 
by which light scattering couples to the microscopic dynamics.}
  \label{Fnorm}
\end{figure}

The whole scenario is compatible with the high-temperature
limit of what has been observed in many glass-forming systems.
We note that the picture does not change up to the highest 
accessible temperatures:
little below the boiling point,
the $\alpha$ peak is still separated by almost a decade in frequency
from the vibrational shoulder.

\subsection{Absolute Intensities}\label{Sabs}

\begin{figure}
\epsfxsize=0.7\textwidth\centerline{\epsfbox{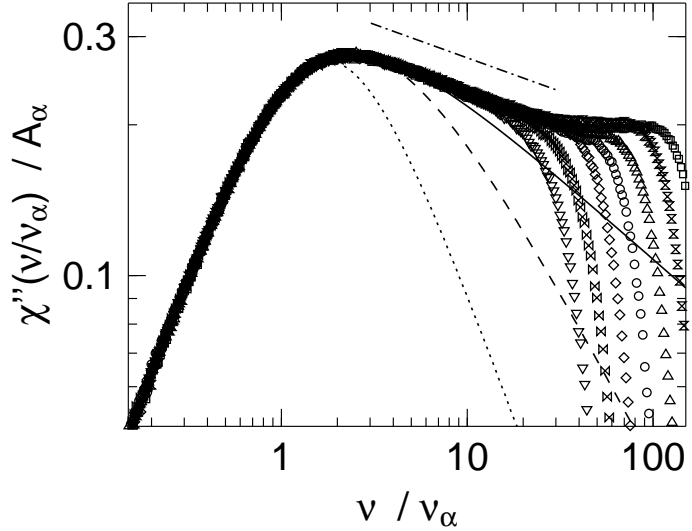}}\medskip
 \caption{Master curve,
constructed by rescaling the light scattering data
of all seven temperatures (same symbols as in Fig.~\protect\ref{Fsusfit})
to a common $\alpha$ peak.
The peak is more stretched than any of the usual fit formul\ae\ 
is able to describe:
a Lorentzian, motivated Maxwell-Debye viscoelastic theory,
is completely inadequate (dotted line).
Kohlrausch's stretched exponential 
(Eq.~\protect\ref{Ekww} with $\beta_{\rm K} = 0.73$,
dashed line)
and the Cole-Davidson function 
(Eq.~\protect\ref{Ecd}, straight line)
hold at least up to above the maximum of the peak.
At higher frequencies,
the extremly flat wing approximately follows a power law
$\nu^{-b}$ with $b \simeq 0.13$ (dash-dotted line).}
  \label{Fskal}
\end{figure}

The temperature dependence of the scattering intensity is surprising:
the height of the $\alpha$ peak increases by about 15~%
on heating from 258 to 293~K;
then it falls back and reaches at 352~K about the starting level.
The apparent anharmonicity in the high-temperature, high-frequency limit
is unexpectedly pronounced,
though a similar trend has been observed in several other liquids.

To disentangle possible causes of these anomalies, 
we take advantage of mode-coupling fits.
The physical meaning of these fits will be discussed later 
 (Sect.~\ref{Sexpmct});
for the moment we take them just as a smooth parametrization
of the measured data --- with one specific advantage:
Since the mode-coupling susceptibilities are obtained by Fourier transform
of the derivative of a time correlation function,
they obey the $\chi''(\nu)/\nu$ sum rule by construction.
Therefore we can take them as representing our light scattering data
in absolute normalization.

Resulting curves are shown in~\ref{Fnorm}:
they are strictly identical to the mode-coupling fits 
included in Fig.~\ref{Fsusfit} 
--- except that the latter are multiplied by an amplitude $A_{\rm MC}$
to adjust them to the arbitrary experimental intensity.
Thanks to the intrinsic normalization,
Fig.~\ref{Fnorm} shows a highly regular temperature dependence.
In particular,
we see no longer indications for a softening of the microscopic excitation
spectrum at high temperatures:
in the high-frequency wing,
up to the boiling point all susceptibilities coincide,
as expected for harmonic motion.
The $\alpha$ peak height increases steadily with $T$;
only between 1 and 2~THz little experimental imperfections are visible.

The temperature dependence of $A_{\rm MC}$ is shown in
the inset of Fig.~\ref{Fnorm}.
Up to 293~K the $A_{\rm MC}(T)$ scatter somehow,
then they decrease systematically towards 
 about 75~\% of the low-temperature average.
A similar decrease of depolarized scattering intensity
has been observed in many other liquids.
However, lacking a means of absolute normalization,
it was never clear whether this decrease reflected a property of the sample
or of the scattering process.
Unstabilities of the experimental setup added to the difficulty.
Fig.~\ref{Fnorm} it appears now
that the decrease of scattering intensities at high temperatures
is {\it not} due to the sample dynamics;
it rather appears that $A_{\rm MC}(T)$ reveals a temperature variation
of the Pockels coefficient that couples light scattering to
the microscopic dynamics.

\subsection{\boldmath{$\alpha$} Relaxation}\label{Sarxexp}

For a quantitative analysis of $\alpha$ relaxation,
we first test time-temperature superposition.
Using the frequency-space representation of Eq.~(\ref{Eama}),
and allowing for a temperature dependent amplitude,
we rescale our data onto a master curve.
Fig.~\ref{Fskal} shows that the line shape 
is independent of temperature 
up to at least five times the peak frequency. 

The $\alpha$ peak is obviously stretched,
as can be seen by comparison to a Debye curve (dotted line).
The data are far better described by one of the empirical expressions
(\ref{Ekww}) or~(\ref{Ecd}).
The Fourier transform of the Kohlrausch 
stretched exponential [Eq.~(\ref{Ekww})]
fits the master curve up to about twice the peak frequency
(dashed line)
with a stretching exponent $\beta_{\rm K}\simeq0.73$.

The Cole--Davidson function (\ref{Ecd})
with $\beta_{\rm CD} \simeq 0.33$
describes the master curve to even higher frequencies (solid line).
Consequently,
we use Cole--Davidson fits to determine the mean relaxation time
\begin{equation}\label{Etau}
\langle \tau \rangle = \int {\rm d}t\, \Phi(t)/\Phi(0) 
  = \beta_{\rm CD} \tau
\,,
\end{equation} 
which in turn is used in the iterative construction of the
master curve.

None of the empirical fit functions is able to fully
describe the extremely flat high-frequency wing of the $\alpha$ peak.
For about one decade,
this wing roughly follows a power law $\nu^{-b}$ 
(dash--dotted line in Fig.~\ref{Fskal}).
The exponent is about $b \simeq 0.13$,
the precise value depending on the choice of the frequency range.
Such a power law is reminiscent of MCT,
though it should not be taken literal as an MCT asymptote
(see Sect.~\ref{Sexpmct} below).

\subsection{\boldmath Relaxation Times}\label{Sarxtim}

In a next step, we investigate the temperature dependence 
of the mean relaxation time $\langle\tau\rangle$,
as determined from the Cole--Davidson fits [Eq.~(\ref{Etau})].

\begin{figure}
\epsfxsize=0.62\textwidth\centerline{\epsfbox{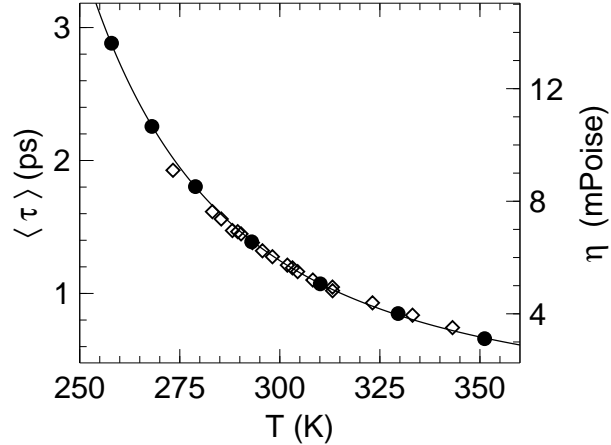}}\medskip
 \caption{Mean relaxation times $\langle \tau \rangle $
as obtained from Cole-Davidson fits to the light-scattering 
susceptibility in the $\alpha$ relaxation region ($\bullet$, left scale).
Solid lines are fits by the Vogel-Fulcher function
(Eq.~\protect\ref{Evft}, Sect.~\protect\ref{Sarxtim}).
The temperature dependence agrees in first order with
that of the shear viscosity $\eta$
($\diamond$, literature data~\protect\cite{And69}, right scale).}
  \label{Ftauabs}
\end{figure}

In Fig.~\ref{Ftauabs} $\langle\tau\rangle$ 
is plotted as function of temperature
and compared to the shear viscosity~$\eta$.
In a good first approximation,
$\langle \tau \rangle$ and $\eta$ show the same temperature dependence.
This completes the demonstration that the susceptibility peak 
under study is indeed due to $\alpha$ relaxation,
in the same sense as in any glass-forming liquid.
Furthermore, the approximate proportionality
$\langle \tau \rangle \propto \eta$ 
can be used to extend available viscosity data \cite{And69}
by more than 15~K into the supercooled phase,
and by 9~K towards the boiling point.

Furthermore, Fig.~\ref{Ftauabs} shows a Vogel--Fulcher fit
\begin{equation}\label{Evft}
\langle \tau \rangle  \propto \exp{\left( -\frac{E_0}{T-T_0} \right)}
\end{equation} 
to the mean relaxation times,
with $T_0 = 94.2 $~K and $E_0 = 650.6 $~K.
We abstain from any physical interpretation,
since several decades in $\langle \tau \rangle$ are needed
for a meaningful verification of~(\ref{Evft});
we just employ the fit as a tool for a more detailed comparison 
of $\langle \tau \rangle$ and~$\eta$.

\begin{figure}
\epsfxsize=0.5\textwidth\centerline{\epsfbox{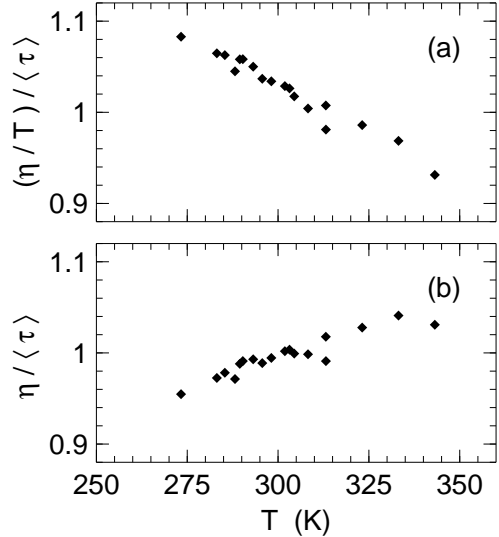}}\medskip
 \caption{For a more detailed comparison of $\eta$ and $\langle\tau\rangle$,
we eliminate their common first-order temperature dependence
by plotting quotients.
Shear viscosity data are the same 
      as in Fig.~\protect\ref{Ftauabs};
the relaxation times $\langle\tau\rangle$ are interpolated to the
  corresponding temperatures by means of the Vogel-Fulcher fit.
The comparison (a) of $\langle\tau\rangle$ and $\eta/T$ is 
motivated by the Stokes-Einstein relation.
However, figure (b) shows that $\langle\tau\rangle$
agrees slightly better with $\eta$ than with $\eta/T$.
Both plots are in arbitrary units.}
  \label{Ftaurel}
\end{figure}

This comparison is performed in Figs.~\ref{Ftaurel}a and~\ref{Ftaurel}b
where we divide either $\eta$ or $\eta/T$ 
by the Vogel-Fulcher estimate of $\langle\tau\rangle$.
A proportionality $\langle\tau\rangle\propto\eta/T$ is suggested
by the Stokes-Einstein relation $D\propto T/\eta$:
when the diffusion constant~$D$ is determined from a time correlation function,
\begin{equation}\label{Ediff}
   \Phi(t)\propto \exp(-Dq^2t)\,,
\end{equation}
one has $\langle \tau \rangle \propto D^{-1}$ and thus 
$\langle \tau \rangle \propto \eta/T$.
Of course 
the stretched $\alpha$ relaxation in benzene
is not correctly described by Eq.~(\ref{Ediff}).
Therefore the theoretical grounds 
for assuming $\langle\tau\rangle \propto \eta/T$
are rather weak.

And indeed, Fig.~\ref{Ftaurel} shows
 that $\langle\tau\rangle$ is not proportional
to $\eta/T$, nor to $\eta$,
but something in between.
Such a temperature dependence has been reported at least once before:
in a neutron spin-echo experiment on the high-temperature dynamics of glycerol
\cite{WuPP96}.
We therefore conclude:
the mean relaxation time observed by scattering  
shows the same temperature dependence as the shear viscosity ---
up to a prefactor of $\cal{O}(T)$ 
which at present no theory is able to predict.

\subsection{\boldmath Mode-Coupling Fits}\label{Sexpmct}

We now extend our analysis beyond the $\alpha$ peak,
considering the full experimental frequency scale up to some THz.
The theoretical reference is given by mode-coupling theory.
MCT is perfectly compatible with the scaling properties
of $\alpha$ relaxation obtained in the two preceeding subsections.
Additionally, MCT predicts that $\alpha$ relaxation has a rather flat 
high-frequency wing which leads over to the
microscopic molecular dynamics.
This is just what we see in Figs.~\ref{Fsusfit}--\ref{Fskal}.

In glass-forming liquids,
mode-coupling analysis usually concentrates on a scaling regime,
designated as fast $\beta$ relaxation,
which is located between $\alpha$ peak and microscopic frequencies.
When the $\alpha$ relaxation becomes sufficiently slow,
the dynamic susceptibility passes through a minimum,
and for frequencies around this minimum simple asymptotic power laws
are predicted.
In benzene we find a susceptibility minimum at the two lowest temperatures.
This once again supports qualitative accord with an MCT scenario,
though the $\beta$ relaxation regime is not sufficiently developed
to allow a scaling analysis.

\begin{figure}
\epsfxsize=0.5\textwidth\centerline{\epsfbox{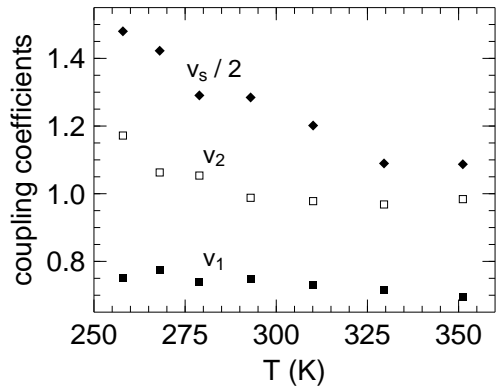}}\medskip
 \caption{
Coupling coefficients used in the schematic mode-coupling fits 
(Figs.~\ref{Fsusfit} and~\ref{Fnorm}, Sect.~\protect\ref{Sexpmct}).
Numeric values are also given in Table~\protect\ref{Tf12}.
The coefficients $v_1$ and $v_2$ of Eq.~(\ref{Ephi})
control the intrinsic dynamics of the liquid,
represented by the correlation function $\Phi(t)$;
the slave correlator $\Phi_{\rm s}(t)$,
which represents the experimental observable,
couples to $\Phi(t)$ via $v_{\rm s}$ [Eq.~(\ref{Ephs})].
In agreement with the spirit of mode-coupling theory,
we find decreasing coefficients with increasing temperature.}
  \label{Fparam-v}
\end{figure}                  

The approximate power law $\nu^{-b}$ (Sect.~\ref{Sarxexp}, Fig.~\ref{Fskal})
in the high-frequency wing of the $\alpha$ peak is 
of the low-frequency asymptote of fast $\beta$ relaxation;
however, the exponent $b\simeq0.13$
implies a line shape parameter $\lambda\simeq0.98$
which, though formally allowed,
is highly unlikely to represent the true asymptotic value,
which whenever reliably determined has been found to fall into
range of about 0.65--0.8.
We therefore think that $\nu^{-b}$ represents not more
than a transient: somehow related to the scaling properties of MCT,
but not representing an analytical asymptote of the $\beta$ minimum.

Therefore, we use numerical instead of asymptotic solutions of MCT.
Specifically, 
we use the two-correlator $F_{12}$ model,
introduced in Sect.~\ref{Smct},
which is defined by the equation of motion (\ref{Eqmo})
[with the set $\{\Phi_q\}$ replaced by the pair $\{\Phi,$ $\Phi_{\rm s}\}$]
and the memory kernels (\ref{Ephi}),~(\ref{Ephs}).
While $\Phi(t)$ models the intrinsic dynamics of the system,
$\Phi_{\rm s}(t)$ shall be interpreted as the correlation function 
observed by depolarized light scattering.

The model contains seven parameters:
two frequencies $\Omega$,~$\Omega_{\rm s}$ characterizing ballistic
 short-time motion,
two damping coefficients $\gamma$,~$\gamma_{\rm s}$ representing
 fast contributions to the memory kernel in Eq.~(\ref{Eqmo}),
and three coupling coefficients $v_1$, $v_2$ and $v_{\rm s}$.
These parameters are all expected to vary smoothly and monotonously
with temperature.
An eighth parameter, the amplitude $A_{\rm MC}$
is not part of the model,
but needed to adjust it to the arbitrary experimental intensity scale;
these amplitudes have already been discussed above 
(Sect.~\ref{Sabs}, Fig.~\ref{Fnorm}).

\begin{figure}
\epsfxsize=0.5\textwidth\centerline{\epsfbox{benz20-8.bb}}\medskip
 \caption{In continuation of Fig.~\protect\ref{Fparam-v},
the remaining parameters of the mode-coupling fits are shown:
(a) the microscopic frequencies 
  $\Omega$ (${\scriptstyle\blacksquare}$) and 
  $\Omega_{\rm s}$ (${\scriptstyle\square}$, kept fixed) 
  of Eq.~\protect\ref{Eqmo}, and
(b) the damping coefficients 
  $\gamma$ (${\scriptstyle\blacklozenge}$) and 
  $\gamma_{\rm s}$ (${\scriptstyle\lozenge}$) representing the
  rapidly decaying part $M_q'(t)$ of the memory kernel.
Since we are using THz units, the figure shows strictly speaking $\Omega/2\pi$
{\it etc.}}
  \label{Fparam2}
\end{figure}

The inner loop of the fit routine calculates $\Phi(t)$ and~$\Phi_{\rm s}(t)$
by iteratively solving Eq.~(\ref{Eqmo}) in the time domain
\cite{Sin95,Got96,Vog98}.
Then $\Phi_{\rm s}(t)$ is converted into a susceptibility
by blockwise Fourier transform, using the Filon method.
The so obtained $\chi''_{\rm s}(\nu)$ are fitted to the experimental data.

\begin{table}
  \label{param}
  \begin{tabular}[th]{llllllll}
 $T$(K)& v$_1$ & v$_2$ & v$_{\rm s}$ & $\Omega/2\pi$ & $\gamma/2\pi$ & 
 $\gamma_{\rm s}/2\pi$ & $A_{\rm MC}$  \\
\hline
258 & 0.7510& 1.172&  2.96&  \phantom{0}821.7& 1283&             2025& 29.07\\
268 & 0.7742& 1.063&  2.845& \phantom{0}892.3& 1337&             1939& 28.31\\
279 & 0.7380& 1.054&  2.581& \phantom{0}803.7& 1086&             2089& 29.78\\
293 & 0.7490& 0.988&  2.569& \phantom{0}905.5& 1027&             1939& 30.09\\
310 & 0.7297& 0.9781& 2.403& \phantom{0}971.4& 1030&             1795& 28.99\\
330 & 0.7159& 0.9682& 2.179& \phantom{0}983.5& \phantom{0}918.1& 1830& 26.54\\
352 & 0.6957& 0.9840& 2.174& 1115&             \phantom{0}785.4& 1674& 22.82\\
\end{tabular}
\bigskip
\caption{Parameters of the two-correlator F$_{12}$~model, 
as obtained from the least-squares fits shown in Fig.~\ref{Fsusfit}.
Note that v$_1$, v$_2$, and v$_{\rm s}$ are dimensionless,
whereas $\Omega$, $\gamma$, and $\gamma_{\rm s}$ are given in GHz.
The amplitude $A_{\rm MC}$ is in the arbitray units of
our depolarized light scattering experiment.
The eighth parameter, $\Omega_{\rm s}/2\pi=1000$~GHz, was kept fixed.}
\label{Tf12}
\end{table}

This procedure is performed independently
for each of the seven measured temperatures.
In an attempt to reduce the number of free parameters
we find that the microsopic frequency of the slave correlator
can be kept at a constant value $\Omega_{\rm s}/2\pi = 1000$~GHz.
All other parameters are found to show a weak, regular temperature dependence
with only minor deviations from monotonicity,
as can be seen in Figs.~\ref{Fparam-v} and~\ref{Fparam2}.
Values are also numerically given in Table~1. 

\begin{figure}
\epsfxsize=0.5\textwidth\centerline{\epsfbox{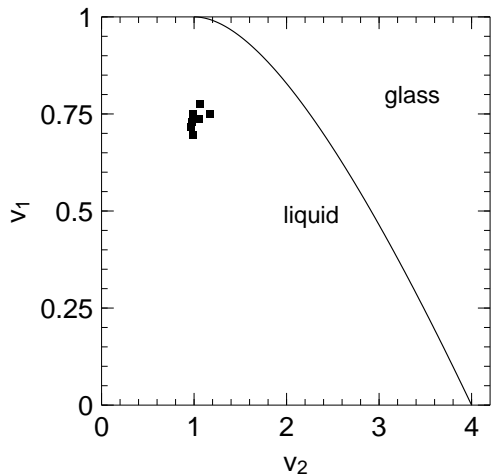}}\medskip
 \caption{Mode-coupling coefficients $v_1$ and $v_2$
as in Fig.~\protect\ref{Fparam-v}
in the phase diagram of the $F_{12}$ model.
The solid line indicates the ideal glass-transition singularity.
On cooling liquid benzene through its entire range of existence,
this phase boundary is only little approached.}
  \label{Fphasdia}
\end{figure}

The time dependence of $\alpha$ relaxation is essentially given
by the coupling coefficients $v_1$ and~$v_2$.
In Fig.~\ref{Fphasdia} the values obtained from our fits are
 shown as points in a phase diagram.
For large values of $v_1$ and~$v_2$,
 the $F_{12}$ model becomes a glass.
The phase boundary
corresponding to the idealized liquid-glass transition 
is indicated in the figure.

In benzene, the coupling coefficients fall clearly in the liquid phase;
with decreasing temperature
the glass-transition singularity is only little approached.
This correlates with the fact that 
the measured susceptibilities show only a very first onset of a
fast $\beta$-relaxation minimum.
For the same reason it would not be meaningful
to use asymptotic scaling laws 
that are based on expansions in $T-T_{\rm c}$.
From the available $v_1$,~$v_2$,
it is not even possible to extrapolate a hypothetical trajectory 
by which benzene would approach the glass transition
if it could be further supercooled.
Therefore it is impossible to indicate a meaningful value
of the asymptotic line shape parameter~$\lambda$.

\section{Conclusion}

We used depolarized light scattering
to measure the dynamic susceptibility of liquid benzene.
Four spectral ranges of two spectrometers
were combined to cover frequencies from 0.5 GHz to several THz.
White noise prevails up to 10 GHz (Fig.~\ref{Fsus}).
Depending on temperature, 
a relaxational maximum is attained between 40 and 180 GHz.
The high frequency wing of this maximum  is 
extremly flat,
and extends up to about 2 THz. 
In the supercooled state, the susceptibility passes
through a slight minimum
around 1~THz (Figs.~\ref{Fsusfit},~\ref{Fnorm}). 

Such a broad relaxation pattern cannot be described by exponential
memory functions that underly conventional theories of simple liquids. 
Instead, our results look very similar to what has been
observed in many glass-forming liquids. 
This confirms our starting hypothesis,
and provides the basis for our quantitative data analysis. 

As in glass-forming liquids, the relaxational $\alpha$~peak
is stretched; it is even more stretched than the common fit formul\ae\
are able to describe. 
Time-temperature scaling
is obeyed with high precision and up to the boiling point,
contradicting certain glass-transition theories which assume
that $\alpha$~relaxation becomes Debye-like 
in the high temperature limit (Fig.~\ref{Fskal}). 

Within the accessible temperature range,
the mean relaxation time $\langle \tau \rangle$ of benzene varies
by more than a factor of 4, 
and it is roughly proportional to the shear viscosty $\eta$
(Fig.~\ref{Ftauabs}).
This accord is \emph{not} improved by 
applying the Stokes-Einstein formula according to which $\langle \tau \rangle$
should go with $\eta / {\rm T}$ rather than $\eta$ (Fig.~\ref{Ftaurel}).
Implications are discussed in Sect.~\ref{Sarxtim}.

Our observations are fully compatible with
mode coupling theory. 
Originally, this theory 
attracted attention because of its ability to model 
a density-driven transition into a nonergodic state. 
Very soon, however, it became clear that this singularity
does not describe glass formation. 
Instead, it is now generally recognized that MCT describes
liquid dynamics at relatively low viscosities. 
In several studies of glass-forming liquids, 
fits were extended above the melting point of the 
concurrent crystalline phase, 
which was found to be irrelevant for the molecular dynamics under study. 
In our present work, 
we push this evolution one step further
by applying MCT to a liquid that can hardly be supercooled 
(actually, benzene \emph{can} be supercooled by nearly 20~K, 
which came out as quite a surprise).
In our experiment, we cover the full range of existence,
up to 1~K below the boiling point. 

In this temperature range, we can no longer apply
the asymptotic expansions that are used in most MCT studies 
of glass-forming materials. 
Instead, we use numeric solutions of the full mode-coupling 
equations of motion.
An elementary model with two correlators and three 
coupling coefficients is sufficient for a satisfactory 
fit to our full experimental data set. 
All parameters show a smooth, 
physically reasonable temperature dependence
(Figs.~\ref{Fparam-v}--\ref{Fparam2}).
Previous mode-coupling studies on glass-forming samples
had mostly concentrated on the asymptotic predictions
for fast $\beta$ relaxation.
A phase diagram makes clear that this scaling regime is
not accessible in benzene (Fig.~\ref{Fphasdia}).
In such a situation numeric solutions of 
a minimal mode-coupling model
provide the most adequate description
of dynamic susceptibilities on
the GHz to THz scale of structural relaxation and microscopic motion 
\cite{x83}.

\section*{Acknowledgments}

We thank M.~Goldammer, W.~G\"otze, H.~Schober, and W.~Petry
for invaluable support.
M.~Fuchs, M.~R.~Mayr, A.~P.~Singh and T.~Voigtmann
showed us how to solve the $F_{12}$ model.
We acknowledge funding by the German DFG under project Me1958/3--1.

\section*{References}

\bibliographystyle{switch-benz}
\bibliography{sw-benz}

\begin{thebibliography}{10}
\bibitem{x74}Empirical formul\ae\ fail to describe the the temperature
  dependence of viscosity simultaneously from the glass transition up to the
  normal liquid phase. Several methods of data reduction indicate that there
  are two different dynamic regimes, separated by $T_{\rm c}$: see Sect.\ 2.3
  in Ref.\ \protect\cite{GoSj92}, and Ref.~\protect\cite{CuLH97a}.
\bibitem{x67}Under the very name ``mode-coupling theory'' the kernel
  (\ref{Ephs}) has already been employed in classical studies of
  tagged-particle motion in simple liquids \protect\cite{BaZo94,WaSj82,MoGl86}.
  However, in those studies the mode-coupling approach was only used for
  $\Phi_s(q,t)$, whereas the $\Phi(q,t)$ to which they couple were typically
  discussed in terms of exponential memory kernels. A self-consistent ansatz
  for $\Phi(t)$ was first proposed in the context of the glass transition
  \protect\cite{BeGS84,Leu84}, and only through the experience with
  glass-forming systems we were led to apply it to the simple liquid benzene.
\bibitem{x85}We have not further analysed the VV spectra. Note, however, that
  we find a nontrivial sound dispersion: Whereas up to some MHz the
  room-temperature sound velocity of benzene is about 1300~m/s, our VV
  Brillouin scattering data give about 1500~m/s at 5~GHz.
\bibitem{x83}Which does not mean that MCT tells the full story about the
  dynamics of benzene. The sound dispersion \protect\cite{x85} for instance
  indicates that there are other, nontrivial relaxation processes on time
  scales below that of $\alpha$ relaxation. It seems plausible that these
  processes have to do with medium-range order; one could imagine that the
  disk-like benzene molecule assemble in form of liquid planes and piles.
  However, these processes are far too slow to interfere with the microscopic
  dynamics we concentrated on.
\end{thebibliography}


\end{document}